\documentclass[aps,prb,reprint,twocolumn,superscriptaddress]{revtex4-1}
\usepackage{amsmath}
\usepackage{graphicx}
\usepackage[pdftex,bookmarks=true,bookmarksopen,bookmarksnumbered,
                colorlinks,
                linkcolor=blue,
                citecolor=blue]{hyperref}
\usepackage{bm}
\usepackage{float}
\usepackage{newfloat}
\usepackage{lineno}
\DeclareFloatingEnvironment[
    fileext=los,
    name=Figure,
    placement=tbhp,
    within=none,
]{suppfig}

\renewcommand{\figurename}{Figure}
\makeatletter
\renewcommand*{\fnum@figure}{{\normalfont\bfseries \figurename~\thefigure}}
\renewcommand*{\@caption@fignum@sep}{ $|$}
\newcommand*{\rom}[1]{\expandafter\@slowromancap\romannumeral #1@}
\makeatother
\newcommand {\ket} [1] {|{ #1 \rangle}}


\newcommand {\ts}  [2] {#1_{\text{#2}}}     
\newcommand {\Si}{$^{29}$Si}
\renewcommand{\vec}[1]{\mathbf{#1}}

\begin{document}


\title{Single-spin qubits in isotopically enriched silicon at low magnetic field}

\author{R.~Zhao}
\thanks{Present address: National Institute of Standards and Technology, 325 Broadway, Boulder, CO 80305, United~States.}
\affiliation{Centre for Quantum Computation \& Communication Technology, School of Electrical Engineering \& Telecommunications, University of New South Wales, Sydney, New South Wales 2052, Australia}

\author{T.~Tanttu}
\affiliation{Centre for Quantum Computation \& Communication Technology, School of Electrical Engineering \& Telecommunications, University of New South Wales, Sydney, New South Wales 2052, Australia}

\author{K.~Y.~Tan}
\affiliation{QCD Labs, QTF Centre of Excellence, Department of Applied Physics, Aalto University, 00076 AALTO, Finland}
\thanks{Present address: IQM Finland Oy, Vaisalantie 6 C, 02130 Espoo, Finland}

\author{B.~Hensen}
\affiliation{Centre for Quantum Computation \& Communication Technology, School of Electrical Engineering \& Telecommunications, University of New South Wales, Sydney, New South Wales 2052, Australia}

\author{K.~W.~Chan}
\affiliation{Centre for Quantum Computation \& Communication Technology, School of Electrical Engineering \& Telecommunications, University of New South Wales, Sydney, New South Wales 2052, Australia}

\author{J.~C.~C.~Hwang}
\thanks{Present address: Research and Prototype Foundry, The University of Sydney, Sydney, NSW 2006, Australia.}
\affiliation{Centre for Quantum Computation \& Communication Technology, School of Electrical Engineering \& Telecommunications, University of New South Wales, Sydney, New South Wales 2052, Australia}

\author{R.~C.~C.~Leon}
\affiliation{Centre for Quantum Computation \& Communication Technology, School of Electrical Engineering \& Telecommunications, University of New South Wales, Sydney, New South Wales 2052, Australia}

\author{C.~H.~Yang}
\affiliation{Centre for Quantum Computation \& Communication Technology, School of Electrical Engineering \& Telecommunications, University of New South Wales, Sydney, New South Wales 2052, Australia}

\author{W.~Gilbert}
\affiliation{Centre for Quantum Computation \& Communication Technology, School of Electrical Engineering \& Telecommunications, University of New South Wales, Sydney, New South Wales 2052, Australia}

\author{F.~E.~Hudson}
\affiliation{Centre for Quantum Computation \& Communication Technology, School of Electrical Engineering \& Telecommunications, University of New South Wales, Sydney, New South Wales 2052, Australia}

\author{K.~M.~Itoh}
\affiliation{School of Fundamental Science and Technology, Keio University, 3-14-1 Hiyoshi, Kohoku-ku, Yokohama 223-8522, Japan}

\author{A.~A.~Kiselev}
\affiliation{HRL Laboratories, LLC, 3011 Malibu Canyon Rd., Malibu, CA 90265, USA}

\author{T.~D.~Ladd}
\affiliation{HRL Laboratories, LLC, 3011 Malibu Canyon Rd., Malibu, CA 90265, USA}

\author{A.~Morello}
\affiliation{Centre for Quantum Computation \& Communication Technology, School of Electrical Engineering \& Telecommunications, University of New South Wales, Sydney, New South Wales 2052, Australia}

\author{A.~Laucht}
\affiliation{Centre for Quantum Computation \& Communication Technology, School of Electrical Engineering \& Telecommunications, University of New South Wales, Sydney, New South Wales 2052, Australia}

\author{A.~S.~Dzurak}
\affiliation{Centre for Quantum Computation \& Communication Technology, School of Electrical Engineering \& Telecommunications, University of New South Wales, Sydney, New South Wales 2052, Australia}

\date{\today}

\begin{abstract}

Single-electron spin qubits employ magnetic fields on the order of 1 Tesla or above to enable quantum state readout via spin-dependent-tunnelling. This requires demanding microwave engineering for coherent spin resonance control and significant on-chip real estate for electron reservoirs, both of which limit the prospects for large scale multi-qubit systems. Alternatively, singlet-triplet (ST) readout enables high-fidelity spin-state measurements in much lower magnetic fields, without the need for reservoirs. Here, we demonstrate low-field operation of metal-oxide-silicon (MOS) quantum dot qubits by combining coherent single-spin control with high-fidelity, single-shot, Pauli-spin-blockade-based ST readout. We discover that the qubits decohere faster at low magnetic fields with $T_{2}^\textmd{Rabi}=18.6$~$\mu$s and $T_2^*=1.4$~$\mu$s at 150~mT. Their coherence is limited by spin flips of residual $^{29}$Si nuclei in the isotopically enriched $^{28}$Si host material, which occur more frequently at lower fields. Our finding indicates that new trade-offs will be required to ensure the frequency stabilization of spin qubits and highlights the importance of isotopic enrichment of device substrates for the realization of a scalable silicon-based quantum processor.
\end{abstract}


\maketitle

\section*{Introduction}

The rapid progress of spin quantum bits (qubits) in silicon quantum dots (QDs) has been fueled by coherent electron spin resonance (ESR), either driven by an ac magnetic-~\cite{Veldhorst2014} or electric-field~\cite{kawakami2014electrical,takeda2016fault}. To date, ESR has enabled the measurement of long coherence times~\cite{tyryshkin2012electron,Muhonen2014,Veldhorst2014,Ladd2018}, and the demonstration of high-fidelity single- and two-qubit logic gates approaching the fault tolerance threshold~\cite{watson2018programmable,zajac2018resonantly,yoneda2018quantum,huang2018fidelity,yang2018silicon,xue2018benchmarking}. In most of these demonstrations, single-shot spin readout was performed via spin-selective tunnelling of a single spin to a reservoir~\cite{elzerman2004single}, however this technique is problematic for the operation of large-scale multi-qubit architectures~\cite{veldhorst2017silicon,jones2018logical} based on industrial manufacturing~\cite{Maurand2016,veldhorst2017silicon,Hutin2018,vandersypen2017interfacing}. In order to work, it would require complex, coherent single-spin shuttling~\cite{baart2016single,fujita2017coherent,nakajima2018coherent} or a chain of swap gate operations~\cite{fogarty2018integrated,sigillito2019coherent} to read out qubits distant from a reservoir. Furthermore, quantum error correction protocols require simultaneous spin-state readout at arbitrary locations in the large-scale qubit array~\cite{jones2018logical}. The preferred method for spin-readout is, therefore, singlet-triplet~(ST) readout via Pauli spin blockade (PSB) in a double-QD system~\cite{Johnson2005}. Besides alleviating the design constraints imposed by reservoir readout, single-shot dispersive ST readout can also leverage state-of-the-art reflectometry technology to enable gate-based sensors with extremely small on-chip footprint~\cite{west2019gate,pakkiam2018single,urdampilleta2018gate,zheng2019rapid,schaal2019cmos}. Furthermore, ST readout forms the foundation of parity measurements, a main ingredient of many quantum error correction protocols~\cite{jones2018logical}, and does not require the Zeeman splitting energy to be much larger than the thermal energy, which is a limiting constraint with reservoir-based readout \cite{elzerman2004single}. It therefore constitutes a robust readout technique that allows qubit operation over a wide range of magnetic fields. Low magnetic fields, in particular, are highly desirable because they enable qubit operation at lower spin resonance frequencies, simplifying microwave engineering and requiring less expensive signal generators. Coherent ESR and high-fidelity ST readout are therefore the key elements for the implementation of a scalable, silicon-based quantum processor~\cite{jones2018logical,veldhorst2017silicon}. Previous experiments have shown the compatibility of these two techniques for electrons in GaAs~\cite{pioro2008electrically,ito2018four}, however in silicon, demonstrations using electron spins have so far remained incoherent \cite{fogarty2018integrated}, while those employing hole spins are yet to achieve single-shot readout \cite{crippa2018gate}.

In this work, we demonstrate coherent, single-electron spin control with high-fidelity ST readout at a dc magnetic field of 150~mT. This new combination of control and readout protocols enables the investigation of the $^{29}$Si nuclear magnetic field seen by a single-electron spin qubit in a 800-ppm isotopically-enriched silicon substrate at low magnetic fields.  We find that the fidelity of ESR control is affected by a drift of the qubit resonance frequency due to fluctuations in the hyperfine field generated by the 800-ppm $^{29}$Si nuclei, and that the amount of this drift increases at lower magnetic field.  The field dependence indicates that the hyperfine field drift is mostly driven by the ESR signal used to measure it, a finding which informs critical engineering trade-offs in how to compensate for nuclear drifts when scaling up silicon spin qubits in the low-magnetic-field paradigm.

\begin{figure*}
	\includegraphics[width=\textwidth]{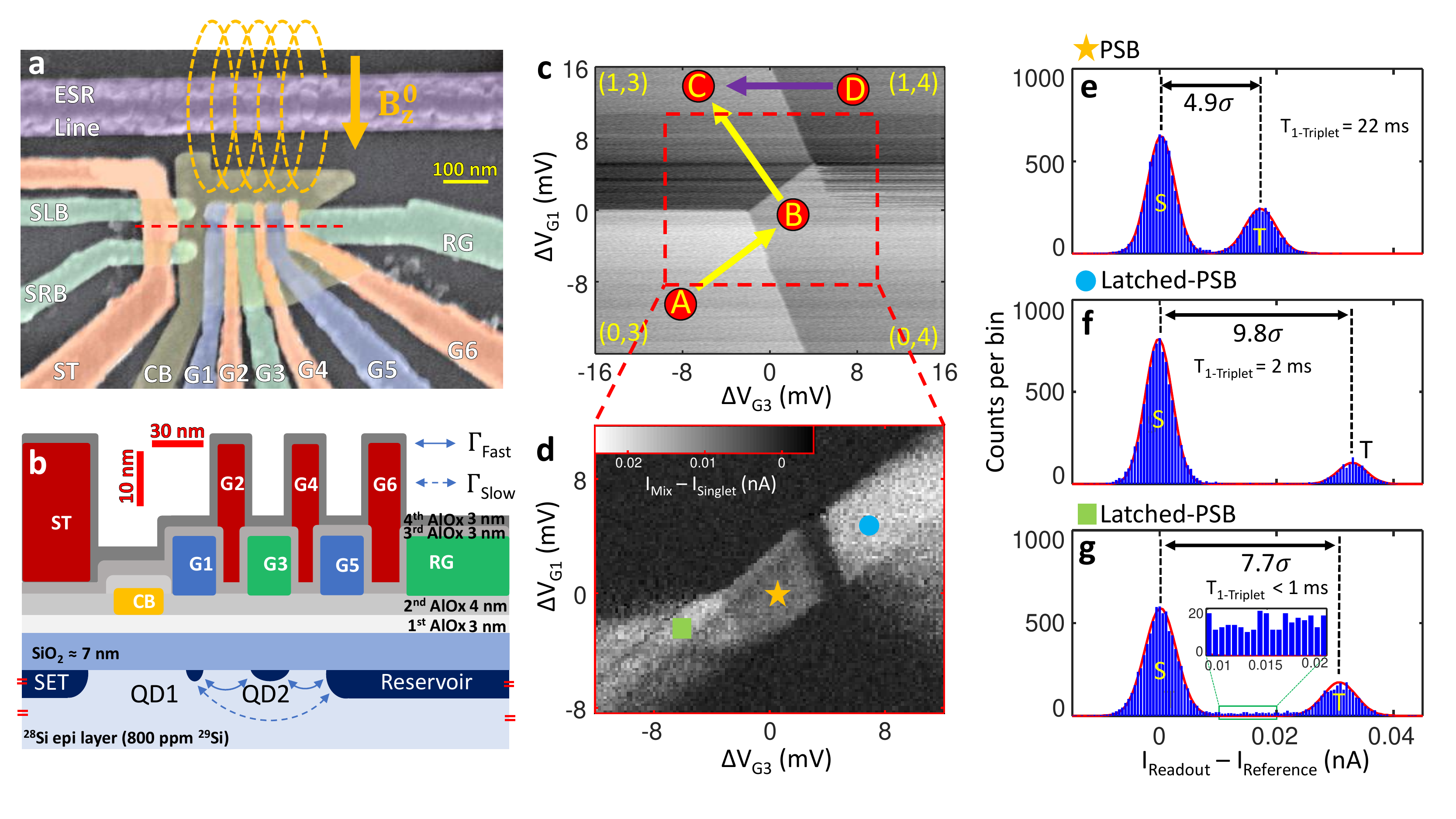}%
	\caption{ \textbf{Device layout and high-fidelity singlet-triplet readout.}
	\textbf{a}, False-coloured scanning electron micrograph of an identical device. The orange arrow indicates the direction of the dc magnetic field $B_\textmd{z}^0$, which is aligned in-plane to minimize spin-orbit coupling and reduce the qubits' sensitivity to charge noise.
	\textbf{b}, Cross-sectional schematic along the red dashed line in \textbf{a}, showing the Pd gate stack with insulating atomic-layer-deposited aluminium oxide layers.
	\textbf{c}, Charge stability map around the (1,3) - (0,4) anti-crossing. The pulse sequence A-B-C (yellow arrows) prepares a separated double dot with electrons in the singlet state, while the pulse sequence D-C (blue arrow) prepares the double dot in a mixed state. The pulse sequence C-B facilitates the spin readout operation by attempting to push the electron from QD1 into QD2. Tunnelling will occur if the electron in QD1 forms a singlet with the electron in QD2, but it is blocked if the electrons form a triplet.
	\textbf{d}, Difference in SET current between singlet and mixed spin state preparation, plotted as a function of the two gate voltages $\Delta V_\textmd{G1}$ and $\Delta V_\textmd{G3}$. Apparent are the regions of standard PSB (star) and enhanced latching readout (circle, square).
	\textbf{e}-\textbf{g}, Histograms of the SET current signal for 10200 single-shot readouts with 120~$\mu$s integration time and $5$~pA bin width, using the \textbf{e}, standard PSB, or \textbf{f},\textbf{g}, enhanced latching readout regions as indicated by the coloured markers in \textbf{d}. Experiments in \textbf{d} - \textbf{g} are performed in a magnetic field of 10 mT.
	\label{fig:Fig1}}
\end{figure*}

\section*{Results}
\subsection*{Device operation and ST readout.}
The device consists of a linear array of silicon metal-oxide-semiconductor (Si-MOS) QDs, electrostatically defined by a palladium (Pd) gate stack~\cite{brauns2018palladium} on an isotopically enriched $^{28}$Si epi-layer~\cite{itoh_watanabe_2014} (see Methods section). Fig.~\ref{fig:Fig1}a shows a false-coloured scanning electron micrograph of a device nominally identical to the one used in this study. A cross-sectional schematic is plotted in Fig.~\ref{fig:Fig1}b. We have labelled the QD accumulation gates (G1-G6), the confinement gate (CB) that confines the QDs laterally, the reservoir gate (RG) that accumulates the electron reservoir for loading and unloading electrons, the single-electron transistor (SET) charge sensor (ST, SLB, SRB), and the short of the broadband ESR antenna to apply MW pulses~\cite{dehollain2012nanoscale} for spin control.

When tuning up the device, we use the following procedure to find the PSB region that gives us high-fidelity, single-shot ST readout at low magnetic field. First, we carefully tune all the gate voltages to accumulate two QDs at the approximate locations indicated in Fig.~\ref{fig:Fig1}b, using the charge sensing technique described in Ref.~\onlinecite{yang2011dynamically}. We then choose the (1,3) - (0,4) anti-crossing, shown in Fig.~\ref{fig:Fig1}c, as our working point. We assume the first two electrons in QD2 to form a singlet and not to interact with the third and fourth electrons during the experiment. Second, we either initialize the two QDs in a singlet state or in a mixed state of singlet and triplet. When pulsing from point A to point B (yellow arrow) we load a fourth electron on QD2, with the last two electrons forming an anti-parallel spin pair (singlet). Further pulsing to point C will separate these two electrons across the two QDs, keeping them in the singlet state \cite{fogarty2018integrated}. Alternatively, when pulsing from point D to C, we will randomly remove one of the four electrons from QD2. The spin of the remaining, third electron could be either anti-parallel or parallel with the spin in QD1, leaving the two QDs initialized in a mix of triplet and singlet. Third, we find the region of PSB, by alternately initializing a singlet and a mixed state, and pulsing to a point around the (1,3) - (0,4) anti-crossing. Fig.~\ref{fig:Fig1}d plots the difference in the SET current for the two different initializations, performed interleaved to reject slow drifts of the charge sensor. When the two QDs are in the singlet state, the single electron in QD1 can tunnel onto QD2 causing a change in charge configuration from (1,3) to (0,4). However, when they are in the triplet state, tunnelling is inhibited by PSB and the system remains in the (1,3) charge configuration~\cite{Johnson2005}. Therefore, initialization into the mixed state will result in some probability for the charge sensor giving a reading that represents a triplet. We detect a clear standard PSB region and two enhanced latching regions, in which the very different tunnelling times from the two QDs to the reservoir are used to cause a state dependent change in total charge~\cite{harvey2018high}.

\begin{table}[b]
	\centering
	\begin{tabular}{|l|l|l|l|l}
		\cline{1-4}
		\rule{0pt}{12pt}$B_\textmd{z}^\textmd{0}$ (mT) &     150              &      300       & 450 &  \\ \cline{1-4}
		\rule{0pt}{12pt}$f_{\textmd{Rabi}}$ (MHz)    &         0.25           &      0.25       &  0.33 &  \\ \cline{1-4}
		\rule{0pt}{12pt}$T_2^*$ ($\mu$s)                &      1.4 $\pm$ 0.21              &    2.5 $\pm$ 0.3       & 3.3 $\pm$ 0.65 &  \\ \cline{1-4}
	\end{tabular}
	\caption{\label{tab:1} Observed $T_2^*$ at three different magnetic fields $B_\textmd{z}^\textmd{0}$ with its corresponding Rabi frequencies $f_{\textmd{Rabi}}$.}
\end{table}

\begin{figure*}
	\includegraphics[width=\textwidth]{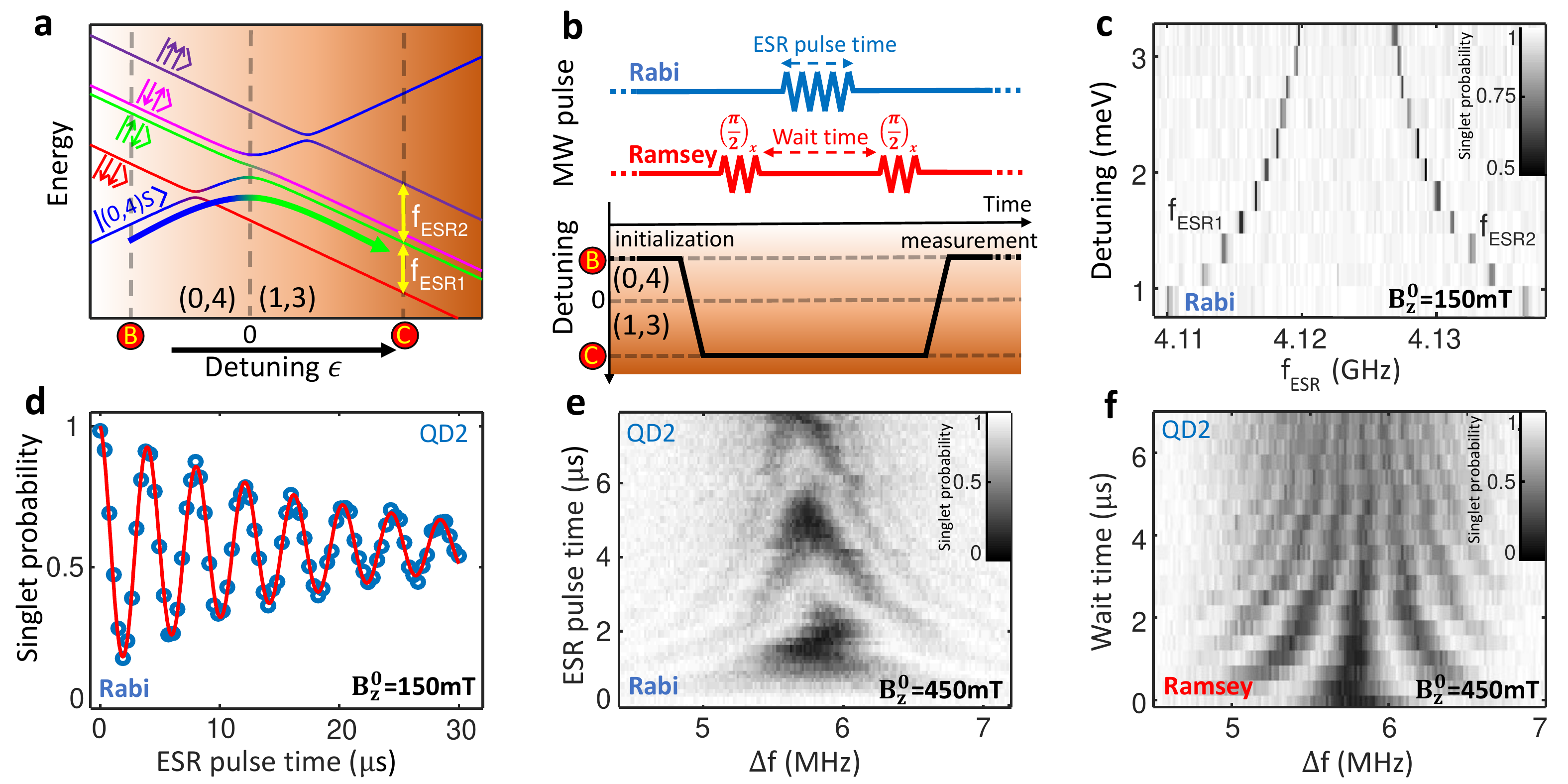}%
	\caption{ \textbf{Coherent single-spin control with ST readout.}
	\textbf{a}, Schematic energy level diagram of the (1,3) - (0,4) anti-crossing.
	\textbf{b}, Partial pulse sequences for the coherent spin control experiments in \textbf{d}-\textbf{f}. The sequences start with initializing the system in the (0,4)S state by pulsing from point A to point B (see Fig.~\ref{fig:Fig1}\textbf{c}). Then we prepare a $\ket{\uparrow\downarrow}$  by moving from point B to point C. As indicated by the curved colour-graded arrow, we pulse diabatically across the S-$\ket{\downarrow\downarrow}$ anti-crossing, but adiabatically with respect to all the other anti-crossings \cite{fogarty2018integrated}. We then apply the microwave control pulses at the deep detuning point C to manipulate the individual spins. After spin manipulation, we perform readout by converting $\ket{\uparrow\downarrow}$ back into singlet and $\ket{\uparrow\uparrow}$ and $\ket{\downarrow\downarrow}$ into triplet, by pulsing from point C back to the standard PSB region B. The $\ket{\downarrow\uparrow}$ state is not accessed during this measurement.
	\textbf{c}, The two ESR peaks plotted as a function of detuning $\epsilon$. We use an incoherent ESR pulse of $100$~$\mu$s for this experiment. We identify the lower (higher) frequency peak corresponding to the QD1 (QD2) through the Stark-shift measurements described in Ref.~\onlinecite{Veldhorst2014}.
	\textbf{d}-\textbf{f}, Coherent spin control of QD2 showing \textbf{d}, Rabi oscillations, \textbf{e}, Rabi chevron, and \textbf{f}, Ramsey fringes. We use frequency mixing to implement the ESR pulse scheme. $\Delta f$ corresponds to the single-sideband modulation of the microwave source. The main carrier frequency of the microwave source is set to 4.2 GHz for \textbf{d} and 12.6 GHz for \textbf{e} and \textbf{f}. We extract the tunnel coupling in the shallow detuning region ($\epsilon \approx 3$ meV) to be $2.0 \pm 0.2$ GHz from \textbf{c}. We performed all the single qubit operations in the deep detuning region ($\epsilon \approx 24$ meV) where the tunnel coupling between the two qubits is negligibly small.
	\label{fig:Fig2}}
\end{figure*}

\begin{table*}[t]
	\centering
	\begin{tabular}{|l|l|l|l|l|l}
		\cline{1-5}
		\rule{0pt}{12pt}Spin state    & Readout outcome & Visibility & Method & Limitation &  \\ \cline{1-5}
		\rule{0pt}{12pt}$\ket{\uparrow\uparrow}$    & Triplet & 87$\pm$1.9\% & ESR driven $\ket{\uparrow\uparrow}\Leftrightarrow\ket{\uparrow\downarrow}$ oscillation & Control &  \\ \cline{1-5}
		\rule{0pt}{12pt}$\ket{\downarrow\uparrow}$  & Triplet & 99.3$\pm$2.3\% & Exchange driven $\ket{\downarrow\uparrow}\Leftrightarrow\ket{\uparrow\downarrow}$ oscillation & Readout  &  \\ \cline{1-5}
		\rule{0pt}{12pt}$\ket{\uparrow\downarrow}$  & Singlet & 99.5$\pm$0.7\% & $\ket{\uparrow\downarrow}\Rightarrow\ket{\downarrow\downarrow}$ $T_{1}$ relaxation & Readout &  \\ \cline{1-5}
		\rule{0pt}{12pt}$\ket{\downarrow\downarrow}$  & Triplet & 96.8$\pm$1.0\% &  $\ket{\uparrow\downarrow}\Rightarrow\ket{\downarrow\downarrow}$ $T_{1}$ relaxation & Thermalization &   \\ \cline{1-5}
	\end{tabular}
	\caption{\label{tab:2} Maximum readout visibility for the four spin states of the two-qubit system. The error bar for $\ket{\uparrow\uparrow}$ state visibility is derived from the fit of ESR driven Rabi oscillations shown in Fig~\ref{fig:Fig2}d. Similarly, the fit of exchange-driven Rabi oscillations, discussed in Methods and shown in Fig.~\ref{fig:ExFig1}c, gives the error bar for $\ket{\downarrow\uparrow}$ state visibility. The exchange driven $\ket{\downarrow\uparrow}\Leftrightarrow\ket{\uparrow\downarrow}$ oscillation experiment is discussed in Methods and reported in Fig.~\ref{fig:ExFig1}. The errors for $\ket{\uparrow\downarrow}$ and $\ket{\downarrow\downarrow}$ state visibility are derived from the fit of the exponential decay curve between the two states as discussed in Methods and shown in Fig.~\ref{fig:ExFig3}}
\end{table*}

\begin{figure*}
    \begin{center}
	\includegraphics[width=0.8\textwidth]{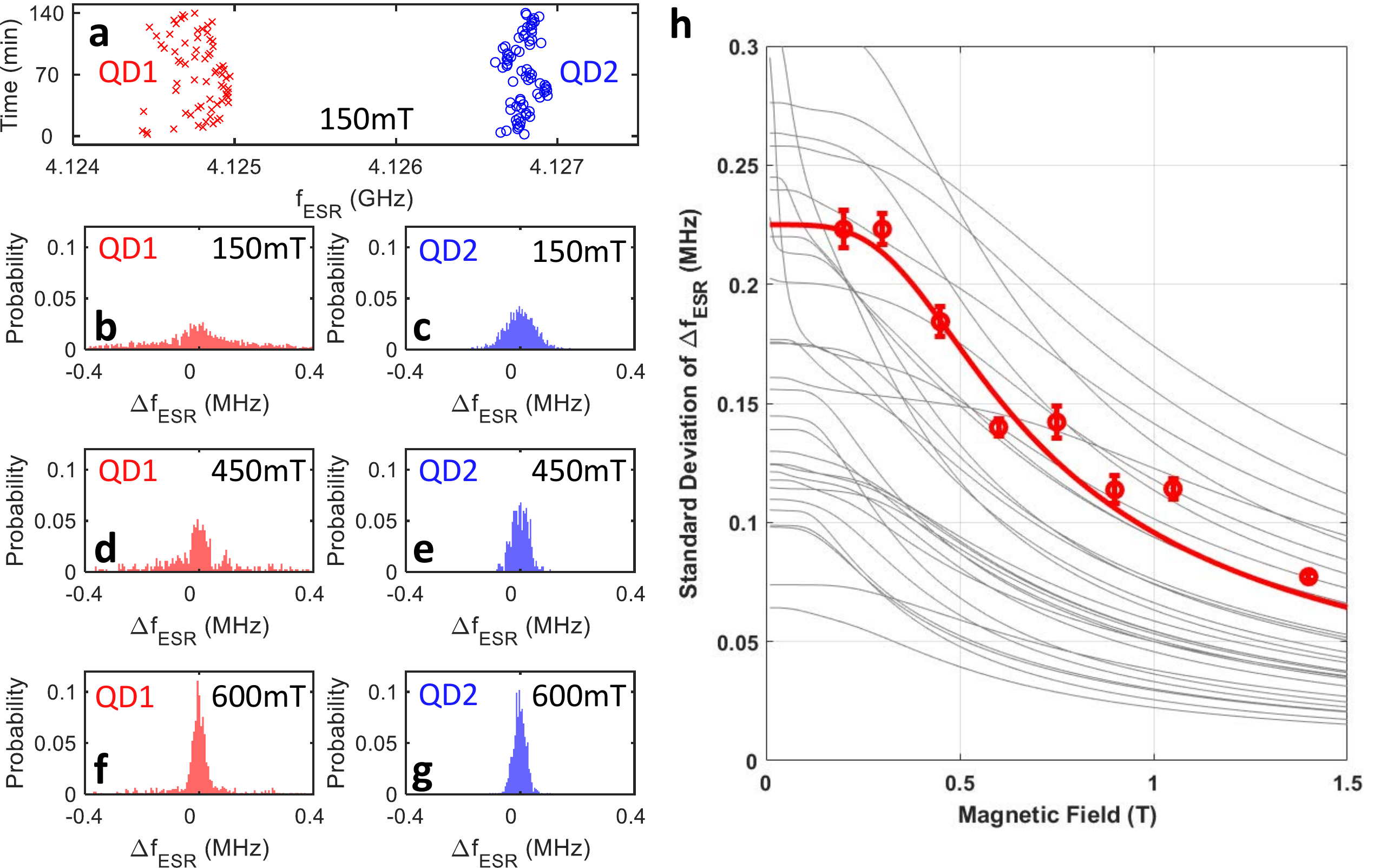}%
    \end{center}
	\caption{ \textbf{$^{29}$Si-induced ESR frequency fluctuations.}
	\textbf{a}, Measurement of ESR frequencies of the two qubits over a time period of 140 mins at 150 mT. There are significant fluctuations of the ESR frequencies of both qubits. Each point of measurement lasts about 2 mins.
	\textbf{b}-\textbf{g}, Histograms of ESR frequency fluctuations for both qubits at various magnetic fields. The distribution becomes more concentrated at higher magnetic field.
	\textbf{h}, Red circles show the standard deviation of the fluctuation in ESR frequency of qubit QD1 plotted as a function of magnetic field, with error bars derived as in Ref.~\onlinecite{rao1973linear}. Each point is calculated based on a measurement that lasts 8 hours.  The red line is a fit going as $1/B_z^0$ at high field and saturating to 0.225~MHz at low field.  Thin gray lines indicate simulations of $^{29}$Si nuclear spin-flip processes for 30 simulated samples with random nuclear placements. The error bars correspond to the standard error of the sample standard deviation estimated from the raw ESR jump data without any assumptions of the shape of the underlying distribution.
	\label{fig:Fig3}}
\end{figure*}

We have designed the orientation of the SET to maximize its sensitivity to inter-dot charge tunnelling events, and achieve good signal contrast for ST readout. As shown in the SET current histogram in Fig.~\ref{fig:Fig1}e, the signal peaks for standard PSB are separated by $4.9$~$\sigma$, indicating a charge state readout fidelity $>99.8$~$\%$. The enhanced latching readout mechanism further increases the signal contrast between the singlet and triplet state, up to $9.8$~$\sigma$ (see Fig.~\ref{fig:Fig1}f,g). Such good contrast would allow charge readout with $<0.005$~$\%$ error, however we find the latched meta state lifetime to be 2~ms or shorter. This is only one order of magnitude slower than the $120$~$\mu$s integration time that is required to achieve good signal separation for our experimental setup. Consequently, we notice a non-zero background between the singlet and triplet peaks in Fig.~\ref{fig:Fig1}g, evidencing decay from the triplet state to the singlet state during measurement. To avoid the spin readout error caused by the latched meta state relaxation as well as mapping errors~\cite{harvey2018high}, we chose the standard PSB region for readout. The triplet lifetime for standard PSB is 22~ms, which results in a $\sim0.5$~$\%$ spin-to-charge conversion error. This reduces the total fidelity of the single-shot ST readout using standard PSB to $99.3$~$\%$. The latched metastate lifetime could be improved in future experiments by defining a third electron under G5 to fine-tune the electron tunnelling rate from QD1 and QD2 to the reservoir~\cite{nakajima2017robust}.

\subsection*{Spin control and coherence times.}
In this study, we control the spins of the electrons using an ac magnetic field generated by the on-chip microwave antenna~\cite{dehollain2012nanoscale}. In Fig.~\ref{fig:Fig2}a we plot a schematic of the energy level diagram as a function of energy detuning $\epsilon$ and in Fig.~\ref{fig:Fig2}b the pulse sequences that we used to achieve ESR. As shown in Fig.~\ref{fig:Fig2}c for $B_z^0=150$~mT, we find two resonance peaks that correspond to rotations of QD1 (lower frequency) and QD2 (higher frequency). When we decrease $\epsilon$, the splitting between the two peaks increases, denoting the larger exchange coupling $J$ between the electrons and mapping out the energy structure of the (1,3) - (0,4) anticrossing in the shallow detuning region. Next, we investigate the coherence of the spin qubits in low magnetic fields. By varying the ESR pulse time, we observe clear Rabi oscillations (Fig.~\ref{fig:Fig2}d) at a magnetic field as low as $B_z^0=150$~mT, with $T_{2}^\textmd{Rabi}=18.6$~$\mu$s. In Fig.~\ref{fig:Fig2}e, we plot the Rabi chevron for one of the qubits at $B_z^0=450$~mT. We also use a Ramsey sequence (Fig.~\ref{fig:Fig2}f) to measure $T_2^*$ at various magnetic fields. The observed $T_2^*$ values are reported in Table~\ref{tab:1}.

The ability to perform coherent control on the spin states allows us to experimentally validate the combined initialization, control and measurement fidelity. We extract the maximum readout visibility for the $\ket{\uparrow\uparrow}$ state by fitting the Rabi oscillations in Fig.~\ref{fig:Fig2}d. The maximum visibility is 87.1$\pm$1.9~\%, which is noticeably lower than our earlier estimate of the readout fidelity. During this measurement campaign, we observed frequent jumps in the ESR frequencies of both qubits due to spin flip of the surrounding $^{29}$Si nuclei. This renders it difficult to keep the microwave frequency exactly on-resonance for the duration of the measurement (see also Fig.~\ref{fig:Fig2}e). Therefore, the true readout visibility of the $\ket{\uparrow\uparrow}$ state may not be reflected in this measurement. We also studied the maximum readout visibility for the three other spin states ($\ket{\downarrow\uparrow}$, $\ket{\uparrow\downarrow}$ and $\ket{\downarrow\downarrow}$) using complementary measurements. The visibility for $\ket{\uparrow\downarrow}$ is as high as 99.5$\pm$0.7 \%, consistent with the readout fidelity estimated previously. The results are summarized in Table~\ref{tab:2} with more details given in the Methods section.

\subsection*{Nuclear spin dynamics.}
The robustness of ST readout allows us to employ tracking of the ESR frequency as a function of magnetic field over a wide range from $B_z^0=150$~mT to 1.4~T.  The existence of a strong field dependence of ESR frequency drifts points to underlying processes arising inside the MOS devices, not compatible with either charge noise or drifts in the applied magnetic field.  As we argue in this section, the residual (800 ppm) $^{29}$Si nuclear spins in the isotopically-enriched substrate are responsible, and their increased role at lower magnetic field is a key result which emerges from the spin control sequences we employ.

As already noticeable in Fig.~\ref{fig:Fig2}e, we observe fluctuations of the ESR frequencies $\Delta f_{\textmd{ESR}}=f_{\textmd{ESR}}(t_2) - f_{\textmd{ESR}}(t_1)$ of both qubits in time. More detailed data is shown in Fig.~\ref{fig:Fig3}a, where we track the ESR frequencies of both qubits over 140~mins. There is no correlation between $\Delta f_{\textmd{ESR}}^{\textmd{QD1}}$ and $\Delta f_{\textmd{ESR}}^{\textmd{QD2}}$, indicating that any fluctuations are caused by a highly localized effect, consistent with a nuclear spin origin, and similar to observations that were published in Ref.~\onlinecite{huang2018fidelity} on a different device. Figures~\ref{fig:Fig3}b-g show the histograms of $\Delta f_\textmd{ESR}$ at three different magnetic fields for both qubits. We observe that the distribution of $\Delta f_{\textmd{ESR}}$ becomes narrower as the magnetic field increases. To better quantify this effect, we plot the standard deviation of $\Delta f_{\textmd{ESR}}^{\textmd{QD1}}$ as a function of magnetic field in Fig.~\ref{fig:Fig3}h.  The increase of this standard deviation at lower magnetic field is not easily explained by nuclear diffusion through dipole-dipole interactions, and suggests other mechanisms for nuclear spin flips.  As these mechanisms provide a critical limiting behaviour for frequency drift and therefore control fidelity, we seek a numerical model to explain it.  The increased fluctuations at lower magnetic field point to the role of the electron spin in the dot, which may more easily drive nuclear spins when the electron and nuclear Zeeman energies are lower.  There are several mechanisms which would show this behavior qualitatively; to isolate which of these are most likely to lead to the observed phenomena, we must look quantitatively at the underlying hyperfine interactions.

\subsubsection{Hyperfine interactions}

First, we simplify hyperfine modeling drastically by treating 2 or 4 electrons as inert relative to nuclei, since overlapping electrons paired into singlet states have no net hyperfine interaction.  Likewise, we treat 3 electrons as a single electron relative to spin interactions, neglecting the other two electrons which form a singlet.   The largest hyperfine interaction for a single electron is the Fermi-contact interaction, with Hamiltonian
\begin{equation}\label{eq:contact}
\ts{H}{en-contact} = \hbar \sum_j A_j \vec{S}\cdot\vec{I}_j,
\end{equation}
where $\vec{S}$ is the single electron vector spin operator and $\vec{I}_j$ is the vector spin operator for the $j$th $^{29}$Si nucleus.  Note that here we put the wafer-normal direction as $x$ and the direction of the in-plane magnetic field as $z$ to conform to common conventions in electron-nuclear magnetic resonance. The principle effect of the contact hyperfine of importance to the present results is the Overhauser shift, given as
\begin{equation}
\delta\omega = \sum_{j} A_j \langle I^z_j \rangle,
\end{equation}
for some configuration of nuclear spin projections $\langle I^z_j \rangle$.

Estimating the size of each $A_j$ requires knowing the shape of the envelope wavefunction $\psi(\vec{r})$ as well as the placement of the $^{29}$Si nuclei.  For the former, we have employed detailed self-consistent Schr\"odinger-Poisson simulations of the device including full three-dimensional gate stacks to capture the potential.  The details of the electrostatic wavefunction model can be found in the Methods section.   We acknowledge that these models do not capture the wavefunction perfectly, as they omit details of electron-electron interactions for closed-shell electrons, have approximate strain models, and do not capture realistic charge offset and disorder effects in real dots.  For the placement of $^{29}$Si, these distribute randomly during epitaxy and so can only be treated statistically.

The contact term is not the only hyperfine term present; we observe the effects of the electron-nuclear dipole-dipole interaction, which is summarized as
\begin{equation}
\label{eq:dipolar}
\ts{H}{en-dipolar} = \sum_j \vec{S}\cdot\vec{D}_j\cdot\vec{I}_j.
\end{equation}
The traceless tensor $\vec{D}_j$, discussed in detail in the Methods section, possesses both diagonal and off-diagonal terms.  The diagonal terms behave similarly to the contact hyperfine interaction, in that a nuclear flip due to this term is always accompanied by an electron flip, therefore costing the electron Zeeman energy $g\ts\mu{B}B_z^0$.  However, these dipole-dipole diagonal terms are nearly two orders of magnitude smaller than the contact terms, as calculated in Ref.~\onlinecite{assali2011}, and so may be neglected.  Much more critical for our model of hyperfine dynamics are off-diagonal, or anisotropic, terms of the coupling tensor $\vec{D}_j$, as these may allow nuclear spin flips without associated electron flips, effectively forming a correction in the local field direction of the nucleus due to the electron spin's polarization.

\subsubsection{Model of driven nuclear fluctuations.}

The dominant term driving nuclear spin flips in the experiments presented here arises from the combination of a tilted local magnetic field for each nucleus due to the anisotropic dipole-dipole interaction in conjunction with the driving of ESR frequencies at the sum or difference frequencies of the electron and nuclear Zeeman frequencies.  This interaction, when driven for times long relative to thermal relaxation times, has been understood for many decades to lead to dynamic nuclear polarization in silicon donor ensembles~\cite{abragam1961}.  In those experiments the process is referred to as the ``solid-state effect."  To summarize the interaction, consider a single nucleus and a single electron.  In the rotating-wave-approximated frame rotating at the applied ESR frequency $\omega$, and keeping only the secular terms of the hyperfine interaction, the Hamiltonian term for~nucleus~$j$~is
\begin{multline}
H_j = (\omega-g\ts\mu{B} B_z^0/\hbar)S^z + \gamma B_z^0 I^z_j + [A_j+D_j^{zz}] S^zI^z_j
\\
 + \Omega S^x + S^z [D_j^{zx} I^x_j+D_j^{zy} I^y_j],
\end{multline}
where $g\approx 2$ is the $g$-factor for electrons in silicon, $\ts{\mu}{B}$ is the Bohr magneton, $\gamma/2\pi=-8.467$~MHz/T is the gyromagnetic ratio for a \Si\ nucleus, and $B_z^0$ is the applied magnetic in-plane field in the $z$ direction.  Because the anisotropic dipolar interaction coefficients $D_j^{zx},D_j^{zy}$ are by far the smallest term in this expression, they may be treated as a perturbation.  In the interaction picture, we find that due to the Rabi drive at rate $\Omega$, the interaction-picture perturbation has oscillatory terms at frequencies $\omega-g\ts\mu{B}B_z^0/\hbar \pm \gamma B_z^0$.  When integrating to first order in the dipolar terms for a Rabi drive of duration $\tau$, at the peak frequencies $\omega=(g\ts\mu{B}/\hbar\pm \gamma)B_z^0$, the first-order probability of a nuclear spin flip is $p_j = (1/32)(\Omega\tau)^2 [(D^{zx}_j)^2+(D^{zy}_j)^2]/(\gamma B_z^0)^2.$ For the experiment with results shown in Fig.~\ref{fig:Fig3}, the pulse duration was $\tau=100$~$\mu$s, and the Rabi drive may be estimated as approximately $\pi/(1~{\mu}s)$, although the Rabi drive was not calibrated in this experiment.  Under these conditions, the peak probability of a flip per Rabi pulse scales as $(1/B_z^0)^2$, varying substantially from one $^{29}$Si to the next, and may be as high as $10^{-4}$ for well coupled nuclei at low magnetic field.  When a flip does happen, the Overhauser shift $\delta\omega$ changes by $A_j$; hence the Overhauser variance for a single nucleus after $N$ trials, treating each nucleus as a balanced random telegraph process, is $[1-(1-2p_j)^{2N}] A_j^2/4$.  We estimate the total variance of observed frequency shifts by summing this variance over all $^{29}$Si nuclei (treated, appropriately, as independent), for a number of trials $N$ given by the number of single shot measurements used at each applied frequency (200) times the number of frequency sweeps in the data (70).  We approximate in this calculation that each sweep over the ESR frequency always hits each resonance once.

The ESR drive is not the only process that causes a change in a nucleus' local transverse field, enabling a spin flip.  We have also considered the probability that the process of the electron tunneling suddenly changing the local magnetic field of a nuclear spin (``ionization shock"), may cause a nuclear flip.  The associated probability of a flip may be estimated as the squared amplitude of the first-order perturbative mixing of two nuclear spin states due to a particular hyperfine term~\cite{fu2003,pla2014}.  For the contact hyperfine term, this probability is approximately
$(1/16) [\hbar A_j/(g\ts{\mu}{B}B_z^0)]^2.$  For the anisotropic dipolar term, this probability is approximately $(1/16)[(D^{zy})^2+(D^{zx})^2]/(\gamma B_z^0\pm A_j)^2.$  These probabilities are lower than the ESR-induced flip probabilities discussed above, but they may occur for every single measurement for every frequency swept in the experiment.

Other mechanisms for flipping nuclei are even smaller.  We also consider the probabilities of nuclear-nuclear flip-flop due to the contact hyperfine term when the dot is occupied by an electron.  We consider the coherent evolution of each nuclear pair in their local hyperfine field due to this effective interaction, and calculate the variance of ESR frequency shifts assuming each nuclear pair is independent.

Notably, all of the nuclear flip probabilities above scale as the inverse square of the applied field, $(1/B_z^0)^2$.  As the probabilities are very small at high $B_z^0$, calculated variances may be simply summed and each are proportional to the associated probability and therefore to $(1/B_z^0)^2$; the root-mean-square (RMS) deviation therefore scales as $1/B_z^0$, as seen in the experimental data.  At very low field, our expression saturates to a total width, which may be regarded as $1/\sqrt{2}\pi T_2^*$ in the ergodic limit, i.e. the expected measure of $T_2^*$ if averaging over the full range of drift.  The question to be answered by our modeling above is whether the amplitude of that variation numerically matches the data.  We find that our model is consistent with the data, if the $1/e$ diameter of the wavefunction is smaller than about 8~nm.  

\begin{figure}
	\includegraphics[width=\columnwidth]{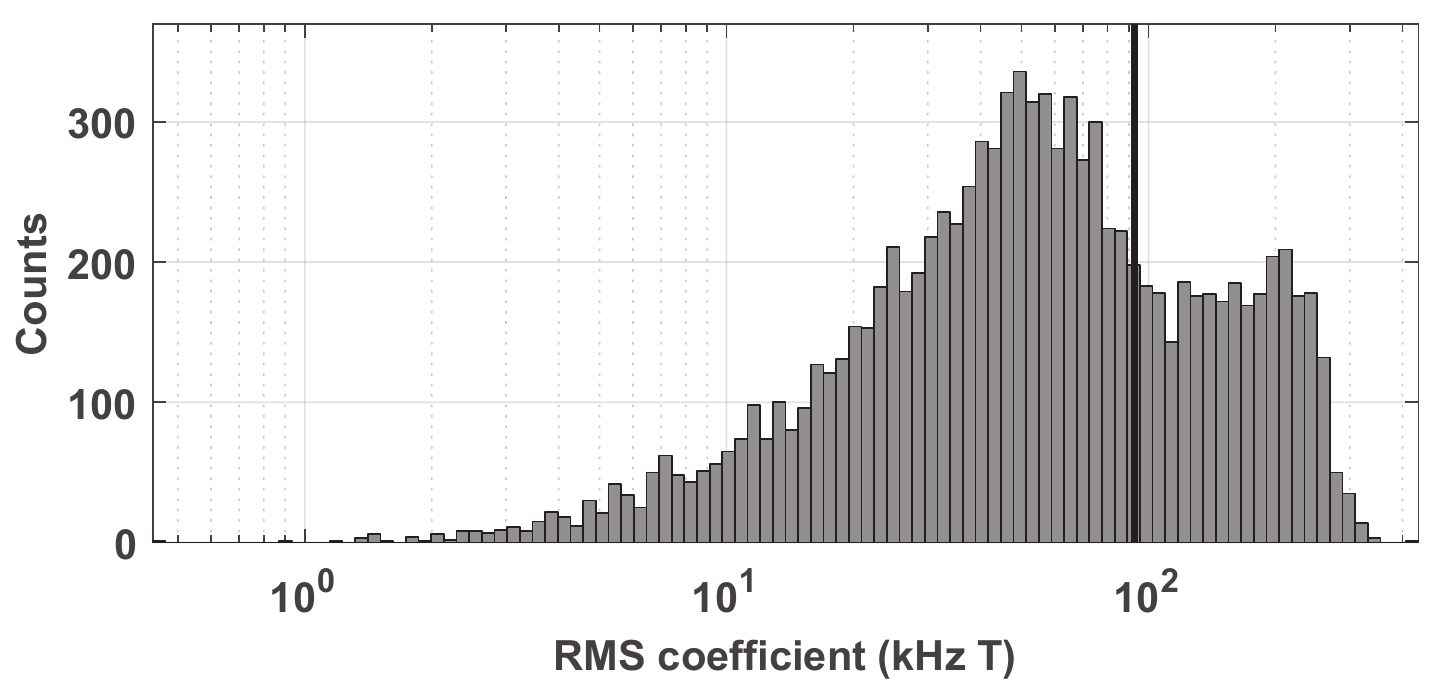}
	\caption{\textbf{ Distribution of RMS deviation coefficients}.  The histogram counts instances for when the fit coefficient $c$, for which the RMS deviation of the ESR frequency varies as $c/B_z^0$, falls within the indicated bin, for a total population of 10,000 random \Si\ placements.  The vertical black line marks the location of the experimentally observed data, indicating a typical population member.}
	\label{fig:Fig4}
\end{figure}

To evaluate the typicality of our real sample, we look at the coefficient for the RMS frequency derivation, $c$, such that $\ts{\Delta f}{ESR} = c/B_z^0$ at high $B_z^0$.  We calculate $c$ for 10,000 virtual samples, each with the same wavefunction size corresponding to vertical field $F=20$~mV/nm and $1/e$ diameter 7~nm, in which the 800~ppm $^{29}$Si are randomly placed relative to the wavefunction.  Figure~\ref{fig:Fig3}h shows example RMS frequency deviations as calculated by finding all probabilities and variances for all $^{29}$Si nuclei as a function of magnetic field for 30 of these virtual sample crystals, all with a diameter of 7~nm.  We find that some simulated samples have a higher variance and ergodic $T_2^*$ than the data, and some lower.  The full distribution of 10,000 coefficients $c$ is extremely broad and the experimentally observed data sits near the mean of the distribution, as shown in Fig.~\ref{fig:Fig4}.  We further find that in our simulation the variance is highly dominated by the ESR-driving term relative to other spin-flip processes we considered.  Averaged over those virtual samples with a coefficient $c$ within a factor of 2 of the experimentally observed coefficient, we find that 99.4\% of the variance results from the ``solid-state effect," i.e. ESR driving on states mixed by the off-diagonal anisotropic dipolar interaction.  The much smaller contribution of the other interactions we considered is shown in Table~\ref{tab:3}.
\begin{table}
	\begin{tabular}{|p{2.5in}|l|}
		\hline
		\raggedleft
		Nuclear-flip process & Contribution \\
		\hline
		\raggedleft
		ESR drive on anisotropically mixed states & 99.36\%\\
		\raggedleft
		Ionization shock from anisotropic dipolar  & 0.41\%\\
		\raggedleft
		Ionization shock from contact terms & 0.16\%\\
		\raggedleft
		Hyperfine-mediated interactions & 0.07\% \\
		\hline
	\end{tabular}
    \caption{ Simulated contributions to ESR variance}
    \label{tab:3}
\end{table}

Although other processes are certainly at play which may influence the drift of the ESR frequency, we find the theory we have presented here to be highly consistent with the observed data, using a model with only physically constrained parameters.  While many approximations were necessary to estimate the wavefunction size and the nuclear spin-flip probabilities, the broad distribution of outcomes provided by the random placement of 800~ppm \Si\ in the small wavefunction means that small deviations from these approximations would not be discernible.  Our model also indicates that the dominant source of the variance of the ESR frequency would be reduced substantially if limiting the scanning of microwaves to avoid the conditions that drive nuclear spin flips.

\section*{Discussion}
In summary, we have successfully operated spin qubits in a silicon platform with a readout scheme showing fidelity higher than 99.3~$\%$ and the ability to work in a broad range of magnetic fields. The ST readout scheme enables coherent control of single spins at a magnetic field of $150$~mT and a resonance frequency as low as 4.1~GHz. Such lower-frequency qubit operation significantly improves the scalability prospects for future silicon quantum processors, but also reveals important magnetic noise effects that become more significant when operating in a low magnetic field environment. Despite using an isotopically enriched $^\textmd{28}$Si substrate, there remain residual $^\textmd{29}$Si nuclear spins located in the vicinity of the qubits, causing fluctuations in the qubit operation frequencies.  We observe the nuclear spin flip rate to increase as the applied external magnetic field decreases, consistent with a numerical model of nuclear spin fluctuations resulting from the ESR drive and anisotropic dipolar interactions. While these $^{29}$Si nuclear spins could potentially be used as a resource for quantum computation~\cite{Hensen2019}, our findings suggest further improvement of the $^{28}$Si isotopic enrichment may be crucial for building a truly scalable spin-based quantum processor.

\section*{Methods}
\subsection*{Measurement setup.}
Figs.~\ref{fig:Fig1}a and b show a scanning electron micrograph and a cross-sectional schematic of the sample used in this study. It is fabricated on top of a 900~nm thick, isotopically enriched $^{28}$Si epi-layer with 800~ppm residual $^{29}$Si on a natural silicon substrate~\cite{itoh_watanabe_2014}. We first grow a 7~nm thick thermal SiO$_2$ plus a 3~nm atomic layer deposition (ALD) aluminium oxide to prevent gate to substrate leakage. Then, we use electron-beam lithography to write the nanoscale gate pattern and perform resist development in pre-cooled (-20~$^\circ$C) Methyl Isobutyl Ketone solution with ultrasound agitation. In the next step, we evaporate Pd onto the chip to form the gate electrodes. Since Pd has a much smaller grain size than aluminium, we routinely achieve gate features as narrow as 12~nm. A 2~nm thin Titanium (Ti) layer is evaporated prior to Pd deposition to enhance the cohesion of the Pd gate onto the oxide. The above lithography process is repeated several times to form the stack of Pd gate electrodes. Since Pd does not have a native oxide, we use ALD to grow 3-4~nm aluminium oxide on top of the gates to provide electrical insulation between different layers. Lastly, the sample is annealed in forming gas at 400~$^\circ$C for 15~mins to repair the damage caused by lithography.

Stanford Research Systems SIM928 modular programmable voltage sources are used to dc-bias all gate electrodes shown in Fig.~\ref{fig:Fig1}a and b. In addition, voltage pulses from an arbitrary waveform generator (Tektronix AWG7122) are applied to G1 and G3 through a voltage combiner. The bandwidth for dc bias (ac pulses) is 30~Hz (80~MHz). We use a nanoscale on-chip integrated 40-GHz antenna to generate the ac magnetic field to drive the electron spin resonance~\cite{dehollain2012nanoscale}. The antenna is powered by a vector microwave source (Agilent E8267D PSG). The SET sensor current is initially amplified with a transimpedance amplifier (Femto DPLCA-200) with 107 V/A amplification. The amplified signal is fed into a JFET voltage amplifier (SIM 910) with gain of 50. This signal is then low-pass filtered at 100 kHz with an analog filter (SIM 965). The amplified and filtered signal is finally readout with a digital oscilloscope (GaGe digitizer OCE838009). The sample is cooled down in an Oxford Instruments, liquid-helium cooled, dilution refrigerator with superconducting magnet. All measurements are performed at the phonon bath temperature of 37~mK.\\\\

\subsection*{$T_1$ relaxation measurement}

\begin{figure*}
	\includegraphics[width=\textwidth]{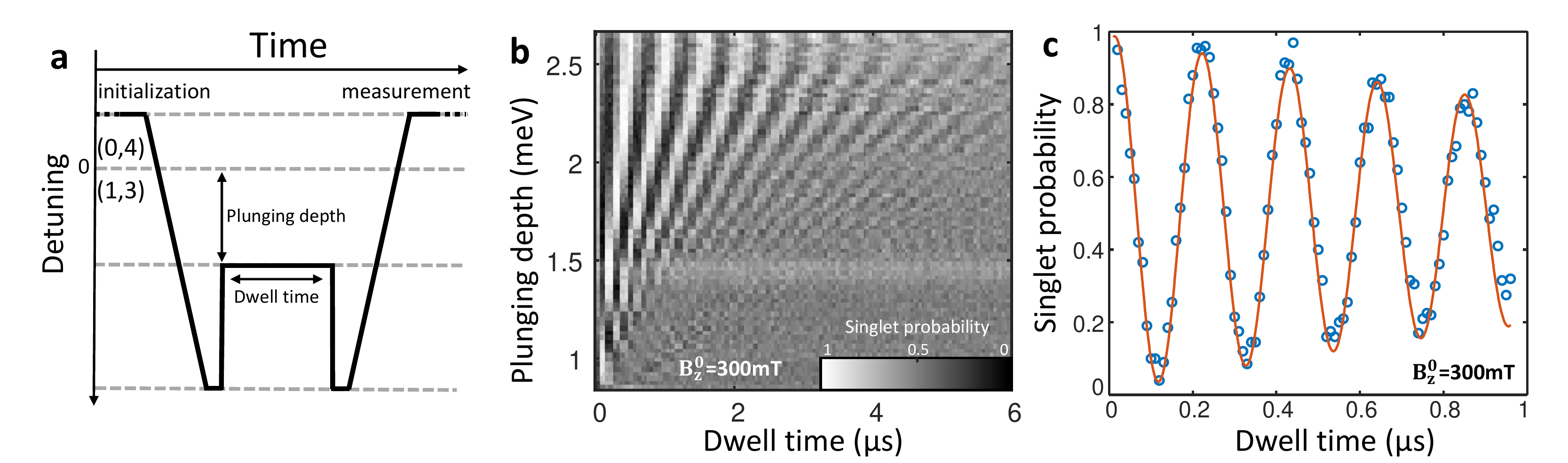}%
	\caption{ \textbf{Exchange driven oscillations.} \textbf{a}, Schematic diagram of the pulse sequence to measure the exchange oscillations. First, we initialize the double quantum dots (DQDs) in the $\ket{\uparrow\downarrow}$ state as described in Fig.~\ref{fig:Fig2}a. Second,we enable the exchange coupling between the two dots by pulsing to the shallow detuning region. This will drive the coherent oscillations between the $\ket{\uparrow\downarrow}$ and $\ket{\downarrow\uparrow}$ states. We then stop the exchange driven oscillations by pulsing back to the deep detuning region. Finally, another adiabatic pulse brings the DQDs from (1,3) back to (0,4) region and completes the spin measurement, during which the $\ket{\uparrow\downarrow}$ state is converted to singlet and the $\ket{\downarrow\uparrow}$ state is converted to triplet.  \textbf{b}, Singlet probability plotted as a function of plunging depth and dwell time at the shallow detuning region.  \textbf{c}, A single trace of exchange driven oscillation with plunging depths of 1.9 meV. The red solid line indicates is a fit to the Rabi`s formula with maximum visibility of 99.3$\pm$2.3 \%. \label{fig:ExFig1}}
\end{figure*}

\begin{figure}
	\includegraphics[width=0.8\columnwidth]{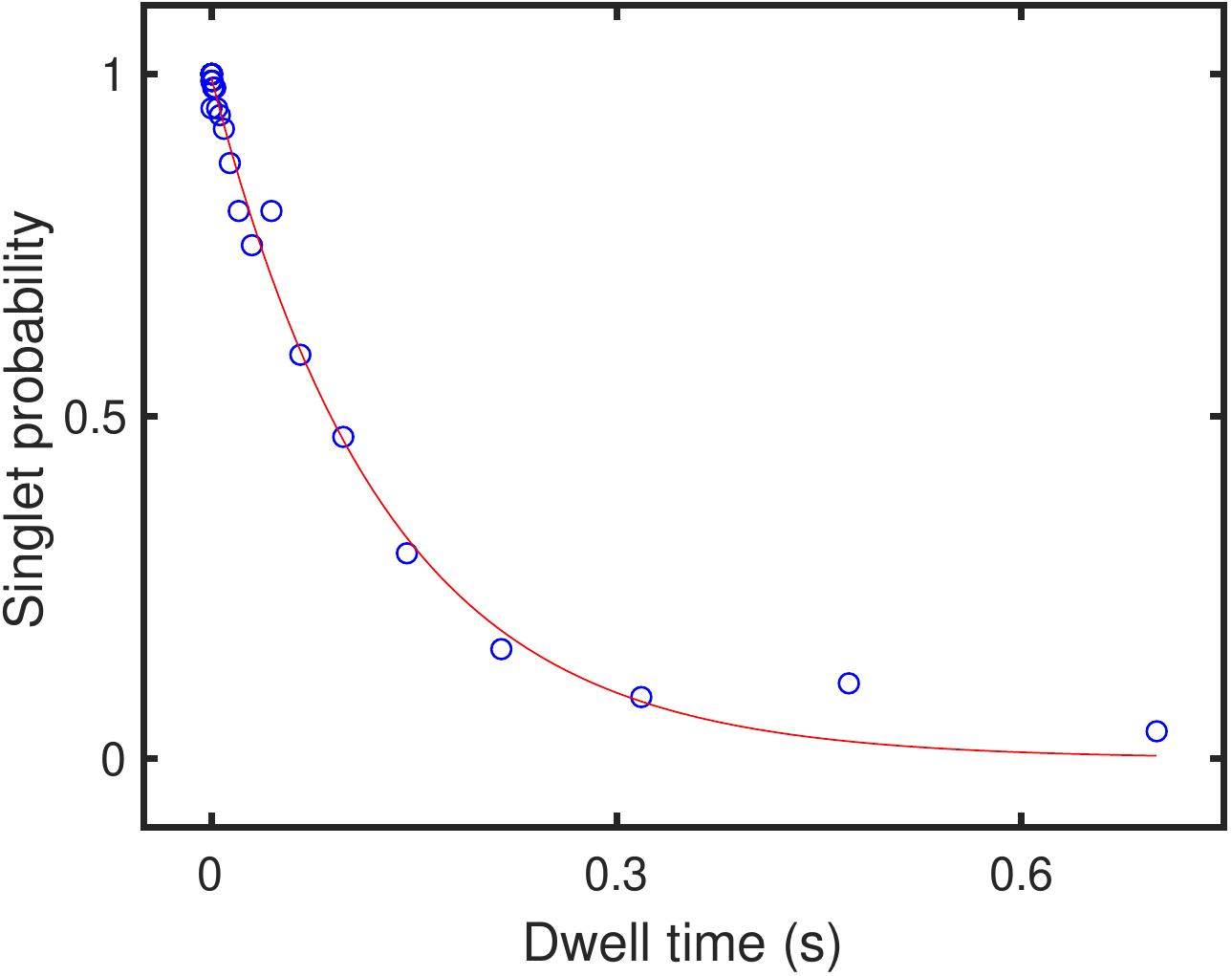}%
	\caption{ \textbf{$\mathbf{T_1}$ relaxation measurement and extraction of maximum visibities of $\ket{\uparrow\downarrow}$ and $\ket{\downarrow\downarrow}$ states.} Decay of the $\ket{\uparrow\downarrow}$ state to the $\ket{\downarrow\downarrow}$ state as a function of dwell time. The red line is a fit to an exponential decay, indicating a relaxation time of 129$\pm$9~ms. Each data point is derived from 200 single-shot singlet-triplet readout events.  All measurements are performed in an external dc magnetic field of 300~mT. We observe the mean of the singlet probability of these eight measurements to be 99.5$\pm$0.7\%. This is the maximum experimentally observed visibility for the $\ket{\uparrow\downarrow}$ state and is consistent with the readout fidelity reported. Similarly, we extract the maximum readout visibility for the $\ket{\downarrow\downarrow}$ state to be 96.8$\pm$1.0\% by setting the dwell time to 0.7~sec. This value could be limited by the 120~mK electron temperature of our system.  We estimate the probability of the $\ket{\downarrow\downarrow}$ state to be thermally excited to the $\ket{\uparrow\downarrow}$ state to be $\ts{P}{excited} = 1 - \exp(-g\ts\mu{B}B_z^0/\ts{k}{B}\ts{T}{electron})=$ 3.4\%, where $g\approx 2$ is the gyromagnetic ratio of the electron, $\ts\mu{B}$ is the Bohr magneton, $\ts{k}{B}$ is the Boltzmann constant and $\ts{T}{electron}$ is the electron temperature of the device.\label{fig:ExFig3}  }
\end{figure}

In this study, we use $T_1$ relaxation from the $\ket{\uparrow\downarrow}$ to the $\ket{\downarrow\downarrow}$ state as an alternative method to assess maximum readout visibilities.  We first initialize the double-quantum-dot (DQD) system in $\ket{\uparrow\downarrow}$ as described in Fig.~\ref{fig:Fig2}a. The DQD system then dwells in the (1,3) region for the $\ket{\uparrow\downarrow}$ state to decay to the $\ket{\downarrow\downarrow}$ state. Finally, we pulse the DQD back to the (0,4) region for spin readout. The $\ket{\uparrow\downarrow}$ state is converted to singlet while the $\ket{\downarrow\downarrow}$ state is converted to triplet. As shown in Fig.~\ref{fig:ExFig3}, the singlet probability decays exponentially with dwell time.

\subsection*{Electrostatic wavefunction model.}

\renewcommand\thesuppfig{\textbf{\arabic{suppfig}}}
\renewcommand{\suppfigname}{\textbf{Extended Data Figure}}
\begin{figure}
	\includegraphics[width=\columnwidth]{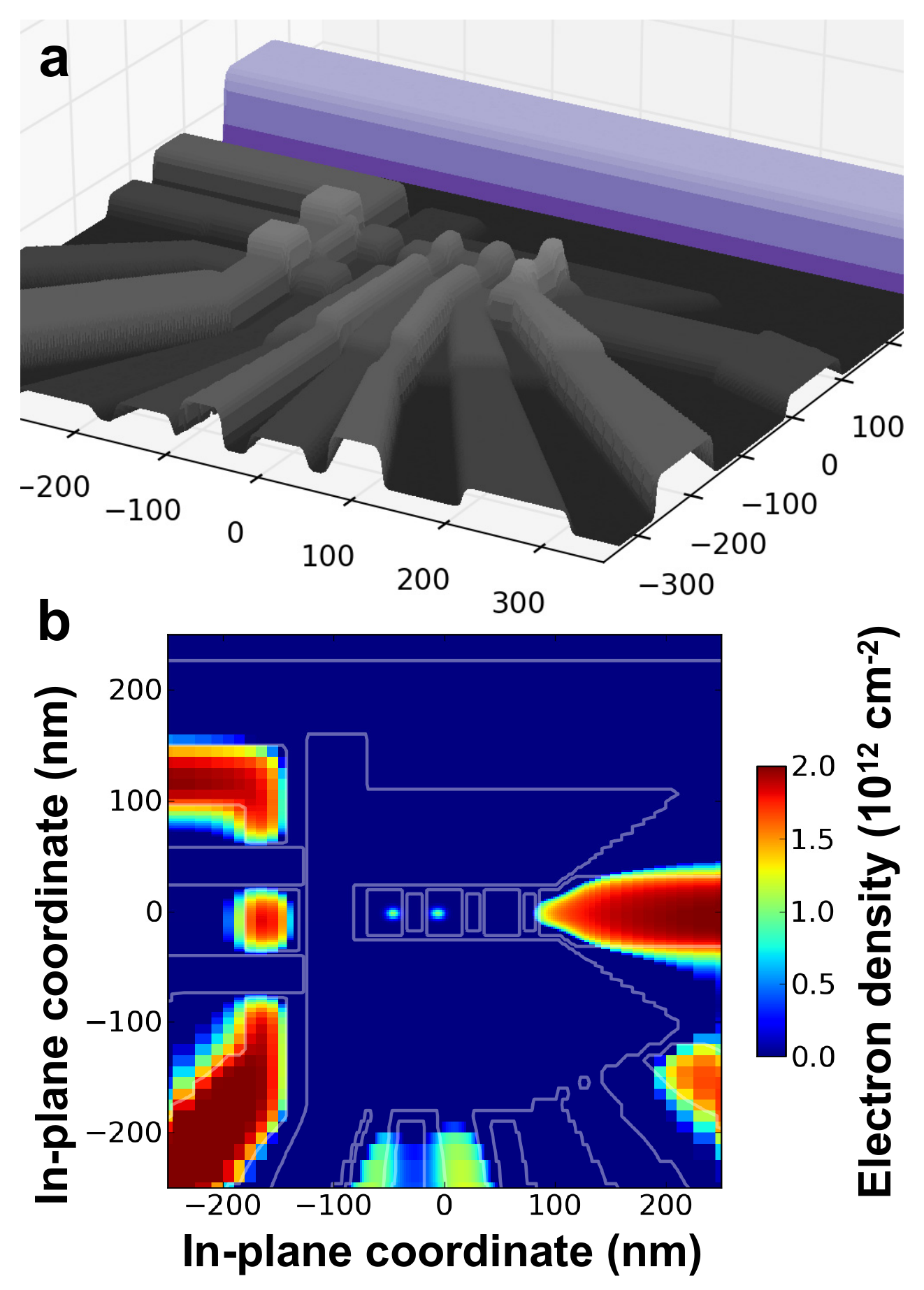}%
	\caption{ \textbf{Self-consistent Schr\"odinger-Poisson simulations of the device. a}, A projection of the full device model layout, showing its ALD-clad top surface (gray) with the ESR line (purple) in the back.  \textbf{b}, Simulated charge accumulation along the Si/SiO$_2$ interface in a nominally tuned device, with the partitioned 2DEG of the sensing SET circuit visible in the left, reservoir puddle in the right, and a pair of interacting electrons in the center, confined by the electrostatically-defined double quantum dot potential. \label{fig:ExFig2}}
\end{figure}

The shape and size of the electron wavefunction defines the degree of overlap with nearby residual $^{29}$Si nuclear spins.  We start with a calculation of the anticipated wavefunction envelope.  To quantify the electron confinement in the inner (quantum) area under the G1 and G3 gates, we have performed self-consistent Schr\"odinger-Poisson simulations of the device. For this, we construct a virtual three-dimensional multilayer and multimaterial device geometry from the production masks and models of the fabrication steps.  An image of the virtual 3D stack appears in Fig.~\ref{fig:ExFig2}a.

Internal stress is expected to build up when cooling this heterogeneous system to cryogenic temperatures.  We have solved the stationary stress-strain problem for the entire layout to incorporate the resulting strain pattern into the electron potential profile as a correction~\cite{thorbeck2015}.  In principle, voids or cracks could form somewhere in the physical device, thus dramatically changing the stress-strain problem, but our simulation treats all constituent layers as perfectly bonded ideal materials.  Another issue is the low reliability and consistency of literature values for elastic parameters of many of the materials in the gate stack; moreover, their applicability down to the scale of tens of (and even few) nanometers is questionable.  However, we find that for this device the strain-induced correction to the potential profile for electrons accumulated at the Si/SiO$_2$ interface is up to a few meV at most, so the limited precision of this calculation is of only modest consequence given the magnitude of voltage biasing applied to the device.

We employ the Thomas-Fermi approximation to model the partitioned 2-dimensional electron gas (2DEG), and full quantum-mechanical treatment of one- and two-electron quantum dot states in the inner area. Figure~\ref{fig:ExFig2}b gives one example of a simulated charge accumulation. When the gates are tuned close to the biases used in the experiments (in which the exchange $J$ between two separated electrons is relatively low), we calculate single-electron quantization energies of $\hbar\omega\sim$16~meV perpendicular to the potential trough and up to $\sim$11~meV along it.  These quantization energies define the spatial extent of the electron eigenstate.  Using $m\approx 0.2m_0$ for the relevant electron effective mass in Si along the oxide interface, the extent $d$ (defined here by the contour of $1/e$ reduction of the electron charge density from its maximal value in the center of the quantum dot) is, approximately, $40\text{~nm} / \sqrt{\hbar\omega\text{~[in~meV]}}$, i.e., the electron state would be about $~10\times12$~nm$^2$ for the quoted quantization energies.  The electron is pressed towards the Si/SiO$_2$ interface by the mean electric field $F\sim$20~mV/nm. These values should be treated cautiously, since our multi-gate MOS device allows substantial flexibility in tuning. Also, in this effort, we have chosen to forego any possible contribution of potential disorder associated with the silicon-oxide interface and/or discrete charges trapped in the oxide.  As such, a smaller wavefunction, as is more likely given the amount of hyperfine fluctuation, is highly plausible.

\begin{figure}
	\includegraphics[width=\columnwidth]{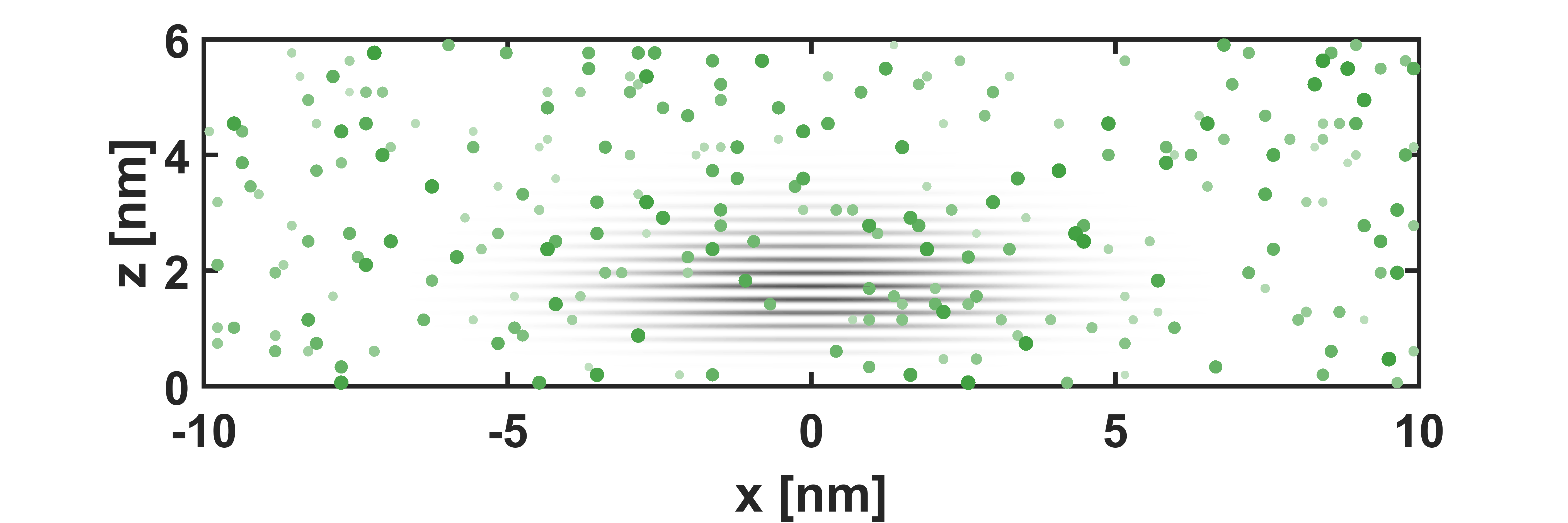}
	\caption{ \textbf{Example nuclear configuration.}  The grey shaded area is an indication of the model for the envelope square wavefunction envelope $|\psi(\vec{r})|^2\cos^2(k_0x+\phi)$. The green dots indicate possible random locations of 800~ppm \Si\ nuclei.
The dot size depicts the proximity of each nucleus to the viewer in this projection.}
	\label{wfsketch}
\end{figure}

In total, our model is consistent with a ``pancake" shaped wavefunction, as anticipated by a simpler model of a vertical triangular potential and parabolic transverse potential.  For the purposes of estimating the wavefuncion to calculate hyperfine coupling parameters $A_j$ and the dipolar coupling tensor $\vec{D}_{j}$, we employ such a simplified model, shown relative to the density of \Si\ for 800~ppm content in a sample nuclear configuration depicted in Fig.~\ref{wfsketch}.  For the particular random configuration shown, 8 nuclei have $A_j/2\pi>30$~kHz.  Another random configuration may have more, or less.  The impact of various hyperfine terms varies considerably depending on where the \Si\ nuclei happened to have landed relative to the small dot and its valley oscillations.

The contact coefficient $A_j$ in units of rad/sec corresponding to Eq.~\ref{eq:contact} is
\begin{equation}
A_j = \frac{2}{3}\mu_0 g\ts\mu{B}\gamma \eta \frac{8}{a^3}|\psi(\vec{r}_j)|^2\cos^2(k_0 x_j+\phi).
\end{equation}
 Here $\eta=178$ \cite{wilson1964} is the Bloch wavefunction overlap, and $a=0.543$~nm is the lattice constant of silicon.  The $\cos^2(k_0 x_j+\phi)$ term arises due to the mixing of 2 equivalent valley states along the growth axis, for which $k_0$ is the wavenumber at the conduction-band minimum and $\phi$ a field-dependent phase offset, and the smooth envelope wavefunction $\psi$ is normalized such that, when summing over \emph{all} lattice sites $r_k$, $\sum_k |\psi(\vec{r}_k)|^2 \cos^2(k_0x_k+\phi) = 1$.

For the dipole-dipole interaction tensor, corresponding to Eq.~\ref{eq:dipolar}, we have
\begin{equation}
\vec{D}_j = \frac{\mu_0}{4\pi} g\ts\mu{B}\gamma \int d^3\vec{r} |\Psi(\vec{r})|^2 \vec{Q}(\vec{r}-\vec{r}_j),
\end{equation}
where the dipole-dipole tensor is
\begin{equation}
Q_{\alpha\beta}(\vec{r})=\frac{3r_\alpha r_\beta-|\vec{r}|^2\delta_{\alpha\beta}}{|\vec{r}|^5},
\end{equation}
$\Psi(\vec{r})$ is the total wavefunction of the electron in the dot, and $\vec{r}_j$ is the location of \Si\ nucleus $j$.  We break the integral into two terms.

The first term ignores envelope variation and focuses on the microscopic integral over the vicinity of the $^{29}$Si in question.  We find that the symmetry of the wavefunction in the valley-mixed state leaves one of the off-diagonal components of the central-cell dipole-dipole coupling tensor non-zero, namely an in-plane $D_j^{\alpha\beta}$ term in which directions $\alpha$ and $\beta$ are along crystalline axes orthogonal to the growth (i.e. valley-mixed) direction.   We evaluate this central-cell off-diagonal term using the plane-wave expansion of the periodic Bloch amplitude, with the numerical coefficients reported in Ref.~\onlinecite{Saraiva2011}.  Handily, integrals are dominated by the volume away from the nucleus, where the expansion is more accurate.  With $\alpha,\beta$ along main crystal axes [100] and [010], this ``on-site" dipolar term is estimated as
\begin{multline}
D_{j,\text{on-site}}^{\alpha\beta} \approx \frac{\mu_0}{4\pi}g\ts\mu{B}\gamma \times 
\\ 
\\0.6 |\psi(\vec{r}_j)|^2 \sin(2k_0 x_j+2\phi) \left(\frac{4}{\sqrt{3}a}\right)^3,
\label{Djzy}
\end{multline}
an expression notably dependent on the valley mixing phase at the nucleus. We caution that this expression is approximate, as microscopic integrals over neighboring cells interfere with the terms we have included.  For the estimate above, the ratio of the central-cell anisotropic off-diagonal term to the contact hyperfine term for a typical nucleus $j$ is a factor of about $10^{-3}$, which is similar to the experimentally confirmed ratio of the same for many $^{29}$Si nuclear sites of a $^{31}$P donor~\cite{ivey1975}, with pronounced exceptions from some of the nuclei very close to the donor.

Although the presence of this term is significant, and must be included in future studies which may allow different magnetic field directions, in the present experiment the magnetic field was 45 degrees different from the crystalline axes, under which conditions this term integrates to zero.  A second, macroscopic term of the dipolar interaction will contribute; we calculate this term by fully neglecting variation in $\Psi(\vec{r})$ over individual atomic cells and so replacing the integral with a sum over all atomic sites excepting the $^{29}$Si in question:
\begin{multline}
\vec{D}_{j,\text{envelope}}\approx 
\\
\frac{\mu_0}{4\pi}g\ts\mu{B}\gamma
\sum_{k\ne j} |\psi(\vec{r}_k)|^2 \cos^2(k_0 x_k + \phi)\vec{Q}(\vec{r}_j-\vec{r}_k).  
\end{multline}
This term is generally nonzero due to finite offsets of the nuclei relative to the MOS electron, and calculated as part of the spin-flip probabilities for the model described in the Results section. We find that the distribution of dipolar term magnitudes has high variation depending on 29Si placement, resulting in the broad distribution of spin-flip rates indicated in Fig.~\ref{fig:Fig4}.



\begin{acknowledgments}
We thank W.~Huang for enlightening discussions and A.~Saraiva for helpful comments on the $^{29}$Si nuclear spin modeling. We acknowledge support from the Australian Research Council (CE170100012), the US Army Research Office (W911NF-17-10198) and the NSW Node of the Australian National Fabrication Facility. The views and conclusions contained in this document are those of the authors and should not be interpreted as representing the official policies, either expressed or implied, of the Army Research Office or the U.S. Government. The U.S. Government is authorized to reproduce and distribute reprints for Government purposes notwithstanding any copyright notation herein. B.H. acknowledges support from the Netherlands Organization for Scientific Research (NWO) through a Rubicon Grant. K.M.I. acknowledges support from a Grant-in-Aid for Scientific Research by MEXT, NanoQuine, FIRST, and the JSPS Core-to-Core Program. K.Y.T. acknowledges support from the Academy of Finland through project No. 308161,314302 and 316551. T.D.L. acknowledges support from the Gordon Godfrey Bequest sabbatical grant
\end{acknowledgments}

\section*{Author Information}
The authors declare no competing financial interests. Readers are welcome to comment on the online version of the paper. Correspondence and request for materials should be addressed to R.Z. (ruichen77@gmail.com) or A.S.D. (a.dzurak@unsw.edu.au).

\bibliography{STqubits}

\end{document}